# THE IN-SITU EXPLORATION OF JUPITER'S RADIATION BELTS

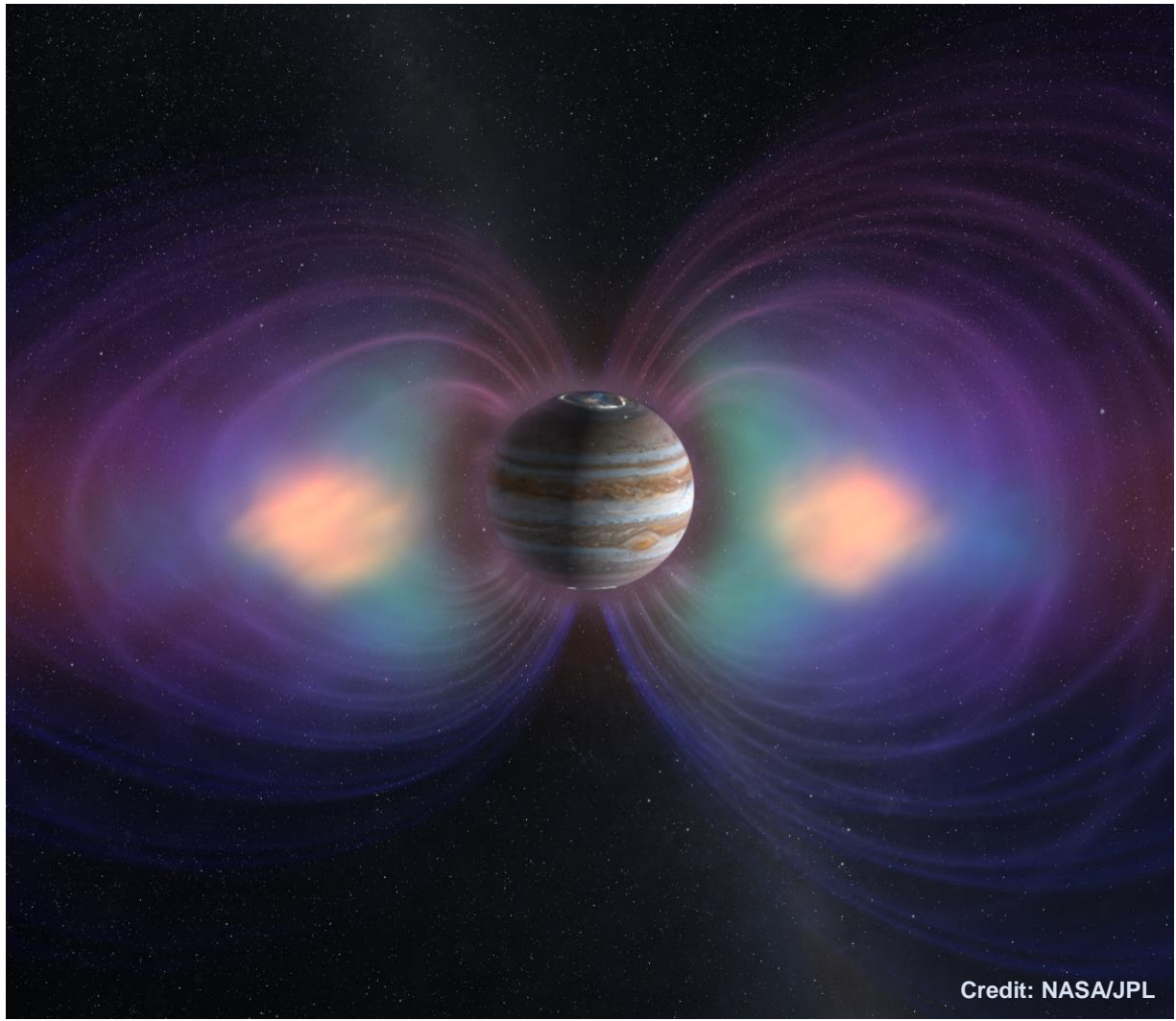

Credit: NASA/JPL

A White Paper submitted in response
to ESA's Voyage 2050 Call

**Contact Person:** Elias Roussos
**Address:** Max Planck Institute for Solar System Research
Justus-Von-Liebig-Weg 3, 37077, Goettingen
Germany
**Email:** roussos@mps.mpg.de
**Telephone:** +49 551 384 979 457

**Executive Summary**

Jupiter has the most energetic and complex radiation belts in our solar system. Their hazardous environment is the reason why so many spacecraft avoid rather than investigate them, and explains how they have kept many of their secrets so well hidden, despite having been studied for decades. In this White Paper we argue why these secrets are worth unveiling. Jupiter's radiation belts and the vast magnetosphere that encloses them constitute an unprecedented physical laboratory, suitable for both interdisciplinary and novel scientific investigations: from studying fundamental high energy plasma physics processes which operate throughout the universe, such as adiabatic charged particle acceleration and nonlinear wave-particle interactions; to exploiting the astrobiological consequences of energetic particle radiation. The in-situ exploration of the uninviting environment of Jupiter's radiation belts present us with many challenges in mission design, science planning, instrumentation and technology development. We address these challenges by reviewing the different options that exist for direct and indirect observation of this unique system. We stress the need for new instruments, the value of synergistic Earth and Jupiter-based remote sensing and in-situ investigations, and the vital importance of multi-spacecraft, in-situ measurements. While simultaneous, multi-point in-situ observations have long become the standard for exploring electromagnetic interactions in the inner solar system, they have never taken place at Jupiter or any strongly magnetized planet besides Earth. We conclude that a dedicated multi-spacecraft mission to Jupiter's radiation belts is an essential and obvious way forward. Besides guaranteeing many discoveries and an outstanding progress in our understanding of planetary radiation belts, it offers a number of opportunities for interdisciplinary science investigations. For all these reasons, the exploration of Jupiter's radiation belts deserves to be given a high priority in ESA's Voyage 2050 programme.

## TABLE OF CONTENTS

# 1 INTRODUCTION & MOTIVATION

## 1.1 Why explore planetary radiation belts?

Radiation belts are the regions of a magnetosphere where high energy charged particles, such as electrons, protons and heavier ions, are trapped in large numbers. All planets in our solar system that are sufficiently magnetized (Earth, Jupiter, Saturn, Uranus and Neptune) host radiation belts *[Mauk and Fox, 2010; Mauk, 2014]*. Radiation belts are not the only regions that high energy particles can be observed; they can be found throughout a planetary magnetosphere, in the heliosphere or the astrospheres of stars, in astrophysical objects such as brown dwarfs, and in the interstellar and intergalactic medium. Many of the environments where energetic particles are found cannot be replicated in the laboratory. Even measuring particle radiation in space is not by itself sufficient to understand its origins: For instance, while we have constrained many properties of Galactic Cosmic Rays (GCRs), the highest energy particles that we can measure, their acceleration sites are inaccessible for in-situ studies.

Radiation belts offer the opportunity to perform ground truth measurements for a variety of high energy physics processes. Apart from containing the energetic particles which we can measure in-situ, they also host most mechanisms that accelerate these particles from low to high energies in a small enough region and over time scales that can be monitored with space missions. These processes are explored in conjunction with additional in-situ particle and fields measurements (plasma, magnetic, electric fields, electromagnetic waves) or close-proximity remote sensing observations, such as Energetic Neutral Atom (ENA) imaging *[e.g. Carbary et al. 2008; Shprits et al. 2012; Roussos et al. 2014]*. These parameters, which are critical for understanding the production and dynamics of particle radiation *[e.g. Woodfield et al. 2013, 2014]*, are similarly challenging to constrain for astrophysical systems. In that sense, planetary radiation belts can be seen as laboratories for in-situ, high energy astrophysics.

The strong links between radiation belts and their host planet further advocate their exploration. Radiation belt particles are modified by the accumulated effects of the planetary neutral environment with which they interact: The properties of planetary exospheres, rings and moon-generated neutral torii are regularly studied through energetic particle measurements, particularly in extraterrestrial systems *[Mauk et al. 2003; Dialynas et al. 2013; Plainaki et al. 2016; Kollmann et al. 2016, 2018a; Nénon and Andre, 2019]*. The reverse path, i.e. the impact of the radiation belts on different components of a planetary system, is also important: Surface sputtering or physical and chemical alteration of moon surfaces are among several fundamental consequences of such an interaction *[Paranicas et al. 2007, 2018; Plainaki et al. 2018; Nordheim et al. 2018]*.

Radiation belt measurements have been performed at all the planets hosting them. The terrestrial radiation belts, studied since the beginning of the space age, are the best understood in terms of structure, origin and dynamical evolution *[Baker and Panasyuk, 2017]*. However, detailed observations of other planetary radiation belts show us that not one of them can be used as a prototype for all others *[Roussos and Kollmann, 2019]*. The need to develop a more universal understanding of how these systems work requires that we establish how radiation belts originate and evolve in the unique space environment of each planet.

In that respect, measurements in the radiation belts of Uranus and Neptune, sampled only once by the Voyager 2 spacecraft, should definitely be part of any future attempt to explore the two planets *[Fletcher et al. 2019]*. Saturn's radiation belts were surveyed in depth thanks to the 13-year Cassini mission at the Kronian system. In comparison, Jupiter's radiation belts, while visited by numerous missions and monitored for decades through their synchrotron emission *[e.g. Han et al. 2018]*, remain largely unexplored. No single mission, payload or observation campaign was ever designed to capture and/or cope with their scale, complexity, dynamics and energetics, as argued in the two follow-up subsections.

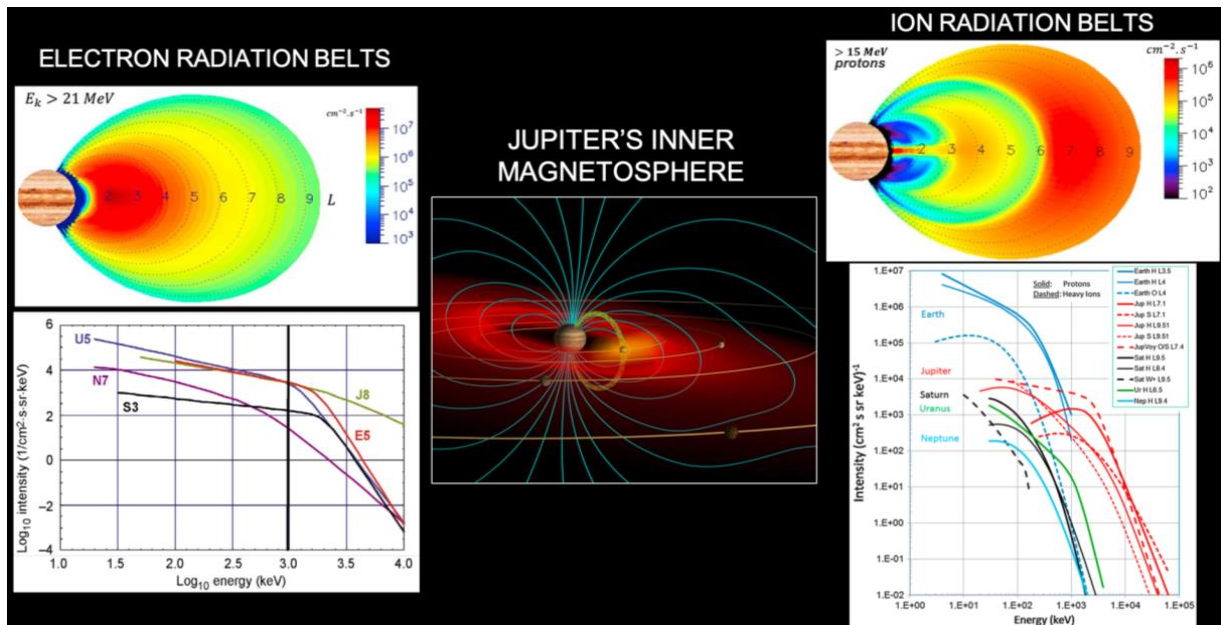

**Figure 1:** Jupiter's magnetospheric region hosting the inner radiation belts (center). The moons Io, Europa, Ganymede and Callisto are drawn, while the Io plasma torus and associated plasma disk are shown in red *(Image Credit: John Spencer)*. Information on the inner electron and ion radiation belts are shown on each side. Color maps are from models of *Nénon et al. [2017, 2018a]*, the synthetic spectra from *Mauk and Fox [2010]* and *Mauk [2014]*

## 1.2   The uniqueness of Jupiter's radiation belts

Jupiter's radiation belts are contained within the planet's magnetosphere, formed by a magnetic field that is 20,000 times stronger than Earth's. Jupiter's fast rotation and material from Io's volcanoes that fills the system, aid the magnetic field to push against the solar wind even further, giving to the magnetosphere enormous dimensions *[Bagenal, 2007]*. Within this giant system, the radiation belts grow into one of the most hazardous regions of our solar system, trapping charged particles of extreme energies and fluxes. Unlike the radiation belts of Earth and Saturn that are limited in their extent *[Ganushkina et al. 2011; Roussos et al. 2014]*, substantial fluxes of energetic particles fill Jupiter's magnetosphere until the magnetopause *[Kollmann et al. 2018b]*. Energetic electrons leaking into the solar wind are so intense that they overwhelm the <10 MeV GCR electrons inside ~10 AU from the Sun, despite Jupiter being a point source in the vast heliosphere *[Potgieter and Nndanganeni, 2018]*.

Jupiter's magnetic field is so strong that even ultrarelativistic, ~100 GeV protons can be trapped near the planet, over 50 times higher in energy than at Earth *[Birmingham 1982; Selesnick et al. 2007]*. Most importantly, observations and theory dictate that processes which may populate the radiation belts with ultrarelativistic particles do exist: Jupiter's inner radiation belts contain electrons with energies in excess of 50 MeV *[Bolton et al., 2002; dePater and Dunn, 2003]*, possibly even above 100 MeV based on model predictions *[Nénon et al., 2017]*. These electrons emit intense synchrotron radiation which can be detected with radio telescopes. This is one of Jupiter's most useful and unique qualities from an observational perspective, since a global picture of the most intense part of the belts can be seen remotely. Data from the Juno spacecraft, currently orbiting the planet, reveal sites at locations remote from the radiation belts, where electron acceleration to MeV energies is continuous and impulsive *[Mauk et al. 2017a; Clark et al. 2018; Paranicas et al. 2018; Bonfond et al. 2018]*.

Jupiter's belts have a distinctively large variety of ions in comparable abundance to protons *[e.g. Garrard et al. 1996; Anglin et al. 1997]*. At other planets, ions at many MeV/n are typically trace elements. Heavy energetic ions at Jupiter, such as oxygen and sulphur, originate primarily from the moons through volcanism or particle sputtering. At lower abundances, ions like helium, sodium, magnesium, carbon etc., have been measured. Furthermore, some of the ions have a range of charge states [*Selesnick and Cohen, 2009; Clark et al. 2016; Dougherty et al. 2017*]. This zoo of particle species and charge states evolving across very broad energy ranges, plasma, magnetic field and wave environments within the magnetosphere, render Jupiter into an unparalleled physics laboratory for testing theories of charged particle transport and non-adiabatic acceleration [*Nénon et al., 2017, 2018a; Moeckel et al. 2019*].

Jupiter's radiation belts are peppered with moons and rings which sculpt them by absorbing and by obstructing the transport of energetic particles. Energetic particle scattering into the atmosphere through waves generated by a moon's electrodynamic interaction with the magnetosphere, as also seen at Saturn [*Santolik et al. 2011*], can have the same effect. Under certain conditions, moons may instead drive charged particle acceleration, mimicking the coupling that may exist between exoplanets and their host astrosphere [*Shprits et al. 2018a*]. The presence of moons within the radiation belts is thus all the more interesting. Studying how the perturbations they generate evolve into the radiation belts and why they seem to affect certain species and energies more than others, we can gain valuable insights into their complex dynamics [*Hess et al. 1974; Selesnick and Cohen, 2009; Nénon et al. 2018a*]. Radiation belt measurements probing physical properties of Jupiter's rings, that are poorly defined by other means, is another important application [*Nénon et al. 2018b*]. Jupiter is also considered as the closest analogue for pulsar and brown dwarf magnetospheres that may host radiation belts [*Kennel and Coroniti 1975; Michel 1979, Williams et al. 2015*].

| Space missions | | | |
|---|---|---|---|
| Mission | Type | Time | Notes |
| Pioneer 10/11 | Flyby | 1973-1974 | Several energetic particle detectors, saturation problems for protons, electrons, radiation damage |
| Voyager 1/2 | Flyby | 1979 | |
| Ulysses | Flyby | 1992 | Several energetic particle detectors, many switched off at Jupiter to avoid radiation damage |
| Galileo | Orbiter, Atmosphere Probe | 1996-2003 | Many orbits through the equatorial belts, several energetic particle detectors, data rate and saturation problems, radiation damage. Limited data by the probe |
| Cassini | Flyby | 2000 | Distant flyby, synchrotron belts monitored by radar experiment, distant ENA imaging |
| New Horizons | Flyby | 2007 | Did not cross into the radiation belts' core, only energetic ions below ~1 MeV(/n) |
| Juno | Orbiter | 2016-2021 | Energetic particle detector, relativistic electrons by monitoring noise in cameras and microwave measurements. High latitude belt crossings. |
| JUICE | Orbiter | 2030-2033 | Mostly >15 $R_J$, energetic particle detectors (<1 MeV), radiation monitor, ENA imagers |
| Europa Clipper | Orbiter | 2026-2029 | >9 $R_J$, dosimeters for high energy particles |
| Other observation modes | | | |
| Type | Example observatories | | Notes |
| Synchrotron Emissions, X-rays | LOFAR, GRMT, VLA | | <50 MeV electrons |
| Aurora (UV, IR, X-rays) | Hubble, XMM, Chandra | | Monitoring energetic electrons (<1 MeV), Heavy ions |
| Io torus remote sensing (UV, X-rays) | HISAKI | | Monitoring large-scale flows, Io volcanism, torus composition |

**Table 1**: List of past, ongoing and future missions to Jupiter's magnetosphere and radiation belts. Examples of remote sensing observing jovian system parameters with direct or indirect relevance to the radiation belts (not exhaustive list). The capability for electron radiation belt monitoring with X-rays is a recent development [*Numazawa et al. 2019*].

## 1.3 Observational challenges and missing links

From the missions that have been or are being used to gain insights into Jupiter's radiation belts (Table 1), none was actually designed to explore them in a comprehensive way. To minimize radiation exposure, spacecraft orbits or operations were planned in ways that avoid the belts' core region. This region is located inward of Io's orbit, at ~5.9 $R_J$ and at low magnetic latitudes (1 $R_J$ = 71492 km, a jovian radius). Pioneer 10, 11 and Voyager 1 reached deeper than Io, offering just brief snapshots of the belts from their flyby trajectories. Future missions JUICE and Europa Clipper will not reach deeper than Europa's orbit, at ~9.5 $R_J$.

For most spacecraft that passed through the inner radiation belts, their instruments suffered from saturation and radiation damage, rendering many of their data unusable or very challenging to calibrate *[Vogt et al. 1979; Fieseler et al. 2002]*. Juno, the only spacecraft passing through the inner belts repeatedly, has instruments responding to <1 MeV electrons and <10 MeV/n ions *[Mauk et al. 2016]*. At higher energies (e.g. >10 MeV electrons), the presence of particles is deduced from the noise they create on instruments like cameras *[Becker et al. 2017a,b]* and thus lack energy or angular resolution. Energetic particle detectors (e.g. Galileo/EPD, *Williams et al. [1992]*) were similarly constrained in terms of energy resolution above few MeV. Energy-resolved measurements are key for radiation belt studies as demonstrated by many works for Earth and Saturn *[e.g. Shprits et al. 2016; Ozeke et al. 2018; Sun et al. 2019]*. Much of the electron physics at these planets are contained below few MeV. At Jupiter, much higher energies, which are hard to measure, are equally important. The properties of waves (e.g. chorus, Z-mode) and plasma distribution functions, which impact the belts' dynamics, are also poorly constrained, especially close to the planet *[Menietti et al. 2016; Moeckel et al. 2019]*. Table 1 lists Earth-based observation methods that offer direct or indirect information for Jupiter's radiation belts. Synchrotron emissions offer direct views for Jupiter's electron belts, but achieve little in terms of energy resolution and provide no data on MeV ions. Monitoring of the Io plasma torus, a product of the moon's volcanic activity, offers insights about magnetospheric flow fields which control the circulation of radiation belt particles *[Murakami et al. 2016; Han et al. 2018]*, but only where the torus is present.

It is clear that these remote sensing observations capture only a small fraction of the big picture. In-situ (ground-truth) observations of the radiation belts are the best way to study them, link their smallest scales to the largest and understand their structure, origin and dynamics. However, with the limitations in existing or future in-situ measurements, this task cannot be fulfilled and a dedicated mission for the belts' in-situ survey is the way forward. The outstanding science that can be performed in Jupiter's radiation belts, discussed in Section 2, justifies why they deserve to be assigned as a high priority target in ESA's Voyage 2050 programme.

## 2 OUTSTANDING SCIENCE IN JUPITER'S RADIATION BELTS

Scientific investigations at Jupiter's radiation belts are well linked to three of the four overarching themes of ESA's Cosmic Vison programme, which should be relevant also for the Voyage 2050 cycle. These links are traced in Table 2 and exemplified in Sections 2.1-2.6.

| ESA Cosmic Vison 2015-2025 | Relevant overarching scientific questions for Jupiter's radiation belts | Specific science goals (Sections) |
|---|---|---|
| How does the Solar System work? | Why are Jupiter's belts so intense and its magnetosphere such powerful accelerator? | 2.1, 2.2, 2.3, 2.4, 2.5 |
| What are the fundamental physical laws of the Universe? | What can we learn for high energy plasma physics by studying different planetary radiation belts? | 2.1, 2.2, 2.3, 2.4, 2.5 |
| What are the conditions for planet formation and the emergence of life? | What are the astrobiological implications of charged particle radiation in Jupiter's mini solar system and especially Europa? | 2.1, 2.2, 2.3, 2.6 |

**Table 2**: Jupiter radiation belt science and the links to ESA's Cosmic Vision themes

## 2.1 Adiabatic electron heating vs local electron radiation belt sources and losses

**State of the art and open questions:** Despite decades of research on how low energy electrons are accelerated to the very high energies observed in Jupiter's radiation belts, many fundamental questions remain open. Similar to Earth *[Green et al. 2004; Horne et al. 2005; Zong et al. 2009]*, two main modes of acceleration are considered: Adiabatic heating and local acceleration of electrons by nonlinear interactions with electromagnetic waves. The challenges involved in separating the two contributions arise from the fact that they overlap in time, space and energy. In addition, the same processes may induce particle and energy losses. For example, whistler-mode chorus waves can either energize electrons through energy diffusion or scatter them into the atmosphere through pitch angle diffusion. Whether a process acts as a source or a sink depends on the background space environment which defines the energies, pitch angles and regions that resonant interactions occur. Figure 2 summarizes our current view on how certain interactions in Jupiter's electron radiation belts (e.g. scattering) are distributed in L-shell and energy and which magnetospheric mechanism is their driving force.

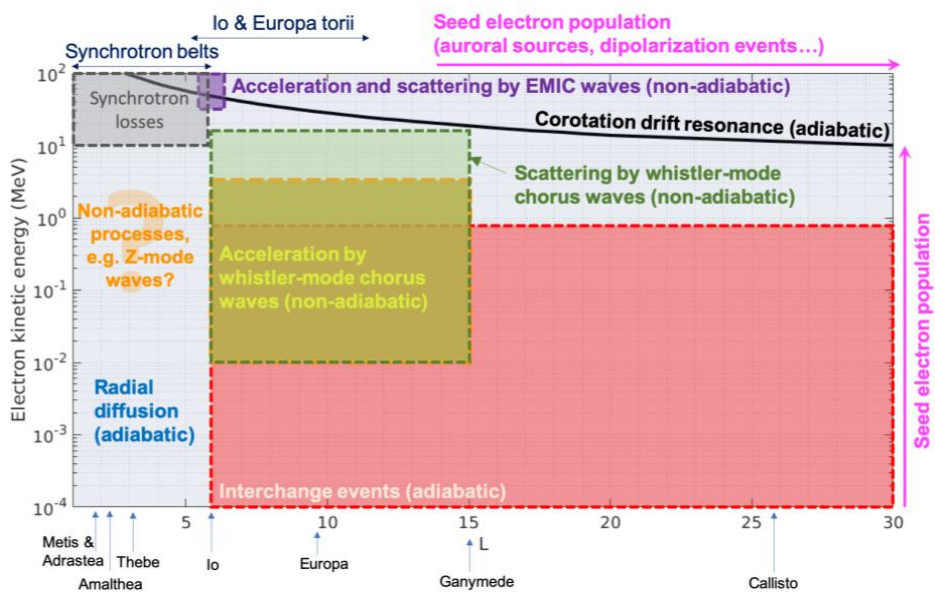

**Figure 2**: A summary processes contributing to electron transport, acceleration and losses in Jupiter's radiation belts as a function of L-shell (magnetic equatorial distance) and energy, as we understand them today. "EMIC" refers to Electromagnetic Ion Cyclotron waves.

Adiabatic heating, i.e. the energization of electrons through inward transport towards stronger magnetic fields, can be facilitated by at least three mechanisms: (1) energy independent radial diffusion induced by magnetic field fluctuations possibly driven by variable thermosphere winds *[Brice and McDonough, 1973; Nénon et al., 2017]*, (2) interchange injections *[Louarn et al., 2014; Dumont et al. 2014]*, and (3) transport due to variable convective electric fields, such as the dawn-dusk electric field *[Murakami et al. 2016]*. The latter refers to a strongly energy dependent mechanism, with high efficiency for 10-100 MeV electrons that drift slowly in magnetospheric local time ("corotation drift resonance") *[Roussos et al., 2018b]*. Due to constraints in existing in-situ measurements, most physical radiation belt models assume an energy independent radial diffusion. Resolving the energetic electron distribution function (e.g. *Sun et al. [2019]*) and the evolution of electron depletions caused by Jupiter's moons at different energies (e.g. *Roussos et al. [2007]*) is necessary for separating overlapping adiabatic processes with a distinct energy dependence (Figure 2). Transient phenomena (e.g. interplanetary shocks) may also mediate adiabatic transport, but are discussed in Section 2.5.

The occurrence of resonant wave-particle interactions has been observed around Jupiter primarily by Galileo in the extended disc of plasma outward of Io's orbit *[Menietti et al., 2016; Shprits et al., 2018a]*. Inward of Io the space environment may also be favourable for wave-driven acceleration, but its properties are less constrained *[Moeckel et al., 2019]*. Other wave types which are not as important at Earth, such as Z-mode, may have a strong impact at the giant planets, as Saturn-based research shows *[Woodfield et al., 2018]*. The radiation belts inward of Io are a prime candidate for Z-mode acceleration of electrons *[Menietti et al., 2012]*. These inner belts are also the only place in the solar system where we can investigate in-situ the impact of synchrotron energy losses. The production of synchrotron radiation not only affects the energy of the electrons but determines how far in latitude they can execute their field-aligned bounce motion, limiting in this way the wave populations they can interact with.

**Key measurements & justification:** Figure 2 shows how any given physical process may generate a seed electron population for another to take over. Adiabatic heating may provide electrons in the appropriate energy range and L-shell where waves would accelerate them further, and vice-versa. Processes thus complement each other and are best studied in unison. In that respect, plasma moments and composition, electromagnetic wave properties (frequency, power, polarization, wave normal angle), energetic electron spectra and their spatial distribution need to be resolved. While Juno, JUICE and Europa Clipper will bridge several existing gaps, low and mid-latitudes and a wide local time range inward of Europa would require coverage. The wave frequency coverage should extend at least until up to the upper-hybrid range, such that the whistler and Z-modes are resolved at any distance, including very strong magnetic field regions near the planet. Electron measurements should be directional and energy-resolved well beyond ~1 MeV. The signature of corotation drift resonant transport could extend up to ~100 MeV, for instance. Observations over extended time should help define the nominal configuration of Jupiter's electron belts.

## 2.2 Cosmic Ray Albedo Neutron Decay as universal proton radiation belt source

**State of the art and open questions:** Galactic Cosmic Rays (GCRs) with a sufficient energy to overcome the barrier of a planetary magnetic field may collide with a planet's atmosphere and rings *[Kotova et al. 2018]*. The secondary neutrons produced by these collisions can travel away from their generation site until they β-decay to protons and electrons. If this happens in a material-free region of the magnetosphere, the decay products become trapped and add to the planet's radiation belts. This process, termed Cosmic Ray Albedo Neutron Decay (CRAND – Figure 3), supplies and maintains the radiation belts of Earth and Saturn with protons of energies between several MeV and up to the proton trapping limit of each planet in the GeV range *[Hess 1959; Cooper et al. 1983; Kollmann et al. 2017a; Roussos et al. 2018a; Selesnick et al. 2018]*. Both Earth and Saturn have moderate magnetic field strengths, that allow a significant part of the GCR spectrum to reach them, and material targets that can generate considerable fluxes of secondary neutrons (Earth's Nitrogen/Oxygen atmosphere, Saturn's icy rings). Jupiter, instead, has a hydrogenous atmosphere, tenuous rings, and a much stronger magnetic field. This different parameter regime is suitable for testing whether CRAND-driven proton radiation belts are a component of any large-scale magnetosphere or if their presence depends strongly on the unique properties of a planet and its magnetosphere.

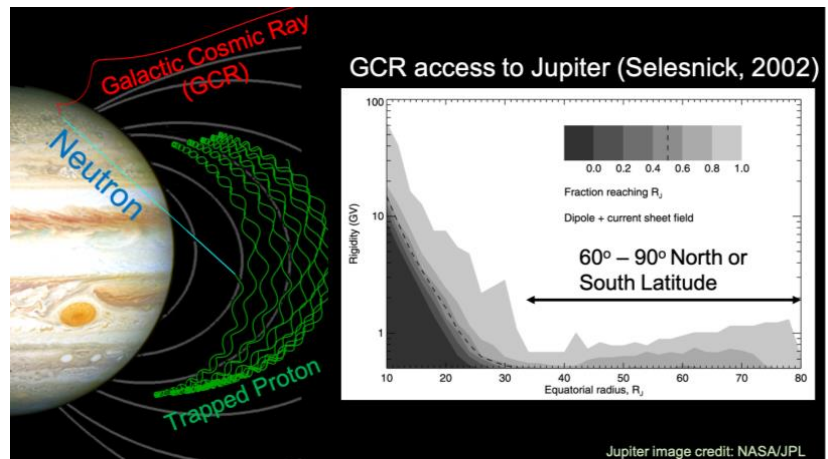

**Figure 3:** The CRAND concept on Jupiter and GCR cutoff rigidities in Jupiter's magnetosphere *[Selesnick 2002]*. The approximate latitude range on the planet (added by the authors) that <1 GV GCRs arrive (0.43 GeV proton), is based on field mapping in *http://www.igpp.ucla.edu/people/mvogt/mapping/*

Existing observations, with proton spectra limited below 80 MeV, do not provide any conclusive evidence: Such protons likely originate through inward adiabatic transport from a distant source [Nénon et al. 2018a]. Jupiter's innermost radiation belts (r<6 $R_J$), where very energetic protons from CRAND may get more easily trapped, are being sampled extensively only by Juno. Juno, however, cannot resolve the energies of protons above 100 MeV, even if the instrument noise they create is recorded [Becker et al. 2017a]. JUICE and Europa Clipper will not reach inward of Europa's orbit and will not provide constraints on CRAND, even if RADEM, the radiation monitor of JUICE, will measure protons up to 250 MeV [Pinto et al. 2019].

From a theoretical perspective, Jupiter's strong magnetic field is expected to exclude a significant part of the GCR spectrum from impacting the planet, thus restricting the production rate of secondary neutrons and the CRAND source to low values [Spergel, 1977]. Still, that same magnetic field may form an efficient trap for CRAND protons compared to the weaker fields of Earth and Saturn. Furthermore, the GCR impact area on Jupiter is very large, while the percentage of neutrons that would decay within Jupiter's enormous magnetosphere before they escape into interplanetary space is higher than at other planets. All that could allow protons to slowly accumulate into a strong proton radiation belt despite a weak CRAND source rate. The efficiency of Jupiter's hydrogenous atmosphere in generating secondary neutrons through processes like proton-proton collisions [Glass et al. 1977] is also unknown. This efficiency has not been possible to constrain at Saturn, Uranus and Neptune, that possess hydrogen-rich atmospheres: atmospheric and ring sources of CRAND at Saturn are mixed and difficult to distinguish [Cooper et al. 2018], while Voyager-2 never reached close enough to Uranus and Neptune where CRAND protons may be stably trapped [e.g. Stone et al. 1989].

**Key measurements & justification:** Simulations predict that magnetospheric (non-CRAND) protons inward of Io drop significantly in intensity above 50-100 MeV and are confined to low magnetic latitudes [Nénon et al. 2018a]. This equatorial confinement becomes stronger inward of Thebe at ~3 $R_J$ (Figure 1, top right). Measurements that extend the energy sampling to the GeV range and with a full pitch angle coverage would resolve CRAND protons directly, if present, given that CRAND is a process that can easily generate protons up to 10s of GeV.

## 2.3   The enigmatic origin of the heavy and light ion radiation belts

**State of the art and open questions:** The jovian radiation belts contain significant fluxes of sulphur and oxygen, resolved up to ~20 MeV/n (Section 1.2 and Figure 1, lower right). The relative abundances of these heavy ions to protons indicate that their origin is primarily iogenic, given the lack of any other obvious sulphur source. As Io generates these species at low energies, there have to be mechanisms accelerating them to the multi-MeV/n range observed. However, the evidence on the nature of the acceleration processes are conflicting. The limited heavy ion charge state measurements for the suprathermal energy range and difficulties to resolve the ion species at many MeV/n with existing data account for that (Figure 4A).

Singly charged energetic ions are most relevant in acceleration models which predict that charge exchange and reionization reactions are important [Barbosa et al. 1984]. These reactions transmit charged particles from the Io torus to an extended region beyond Europa's orbit, which then supplies the ion radiation belts through adiabatic transport. Radial gradients of energetic ion fluxes [Gehrels and Stone, 1983; Cohen et al. 2000] and the abundances of >5 MeV/n sulphur and oxygen [Garrard et al. 1996; Cohen et al. 2001] are consistent with this theory. A recently discovered ion radiation belt component that includes energetic oxygen and sulphur and resides just above Jupiter's atmosphere, also requires that its ions are singly charged [Kollmann et al. 2017b]. Theories that do not involve charge exchange and reionization and set no restrictions on heavy ion charge states, also exist. The energetic ion fluxes beyond Europa's orbit that seed the ion belts can be enhanced following tail reconnection events [Louarn et al. 2014; Kronberg et al. 2019], as at Saturn [Mitchell et al. 2015]. An ion heating source at 20-25 Rj, the nature of which is not understood [Selesnick et al., 2001], may be relevant. Indirect evidence for the presence of multiply charged oxygen and sulphur states come e.g. from *Clark et al. [2016]* through the analysis of interchange injections. *Nénon and Andre [2019]* combine aspects of the aforementioned theories and predict that ionization states evolve with distance: Ions start as multiply charged far from Europa, gradually reducing their ionization through charge exchange as they propagate through the Europa and Io torii. Still, *Selesnick and Cohen [2009]*, who reconstructed signatures of MeV oxygen and sulphur depletions seen by Galileo in the vicinity of Io, concluded that they should be fully ionized (charge states of 8 and 16, respectively). X-ray emissions from the Io torus *[Elsner et al. 2002]* are in part consistent with fully stripped, heavy energetic MeV ions since electron stripping dominates over charge exchange in that energy range. Current estimates of MeV oxygen fluxes, however, seem too low to account for the strong X-ray signal.

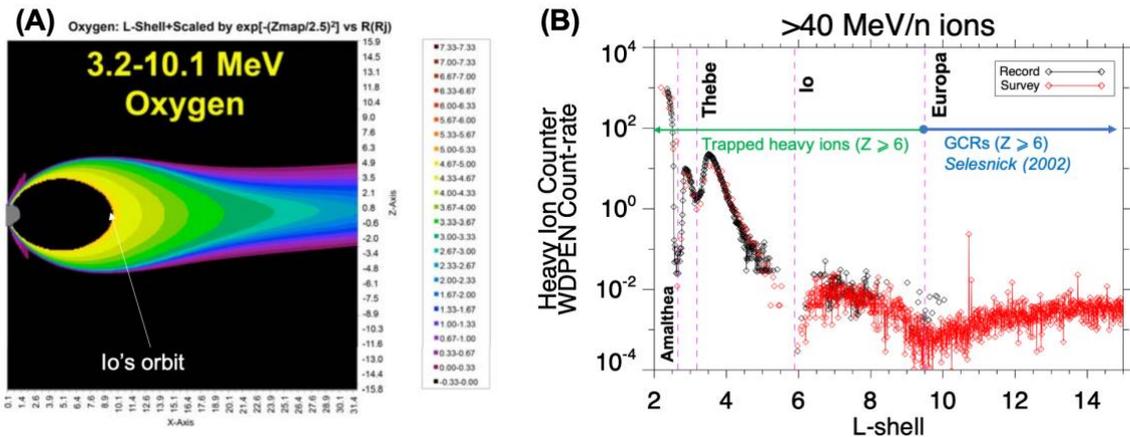

**Figure 4:** (A) Integral fluxes of 3.2-10.1 MeV oxygen at Jupiter, based on an empirical radiation belt model *[Garrett and Evans, 2015]*. The black shaded area inward of Io's orbit is not due to missing data, but to poorly resolved ion-composition (B) Mission averaged L-shell profile of Galileo Heavy Ion Counter count-rates of >40 MeV/n ions, partly published in *Selesnick [2002]* and *Garrard et al. [1996]*. Moon locations are marked, matching well the signal dropouts.

At the highest observed energies, Galileo measurements inward of Io reveal a population of >40 MeV/n and Z≥6 ions which is separated from the rest of the jovian magnetosphere: A strong and persistent signal depletion is seen at Io's orbit (Figure 4B). This implies that for >40 MeV/n heavy ions transport across Io is difficult, if not impossible. With adiabatic transport excluded, the >40 MeV/n species near the planet do not have an obvious source, unless they are supplied across the moon's barrier during extreme episodic events, missed by past spacecraft. If that is not the case, a local production or acceleration mechanism should be invoked. The difficulty to separate species at those energies with existing data means that we cannot exclude that other ions, such as carbon, also contribute. The direct capture and trapping of heavy GCR ions through charge stripping at the upper jovian atmosphere, as seen with Anomalous Cosmic Rays at Earth *[Selesnick et al. 1995]*, may offer a pathway for carbon GCRs into the jovian system. Other energetic ion species that have been detected at Jupiter (sodium, magnesium, helium) can be used as additional tracers of the circulation and acceleration of iogenic or solar wind particles *[Garrard et al. 1996]*, or for the spallation of the jovian rings as a possible source of energetic helium in the radiation belts *[Fischer et al. 1996]*.

**Key measurements & justification:** The heavy ion radiation belts may have multiple components that can be separated by both species and charge state. Each component may be sensitive to different physical processes and its properties (energy spectra, pitch angle distributions) revealing for different aspects of Jupiter's radiation belts and magnetosphere. Many of the existing uncertainties arise from the fact that, for instance, many ion charge state estimations were obtained indirectly. In terms of composition, Juno is filling many gaps for the <10 MeV/n ions *[Haggerty et al. 2017]* but these need to be extended well beyond 100 MeV/n, so as to understand their how their acceleration evolves and what are its limits in energy. Charge states at <1 MeV could still update energetic ion transport and acceleration models, even if they are difficult to obtain for very high energies.

## 2.4 High latitude charged particle acceleration as a universal radiation belt source

**State of the art and open questions:** Jupiter's plasma sheet is the extended closed field line equatorial region beyond Io's orbit that contains dense cold plasma and high fluxes of energetic particles mostly in the keV to low MeV range. The plasma sheet has been studied not only as a major component of the planet's magnetosphere, but also as an enormous storage ring of charged particles that supplies the radiation belts. A large body of research has been dedicated to the dynamics within Jupiter's plasma sheet (such as injections and ion-neutral gas interactions, Sections 2.1 - 2.3), because the majority of the measurements made before the Juno mission were at low latitudes. Despite that, many aspects of its energetics are still unresolved. *Tomás et al. [2004]* and *Mauk and Saur [2007]* observed >30 keV electron angular distributions reminiscent of high latitude auroral processes. Furthermore, Galileo data show the plasma sheet to contain a low but persistent population of >11 MeV electrons, with no obvious origin *[Kollmann et al. 2018b]*. The connection of these equatorial observations to dynamics and phenomena at high latitudes was speculated, but the lack of in-situ data from the magnetospheric high-latitude counterparts left many questions open.

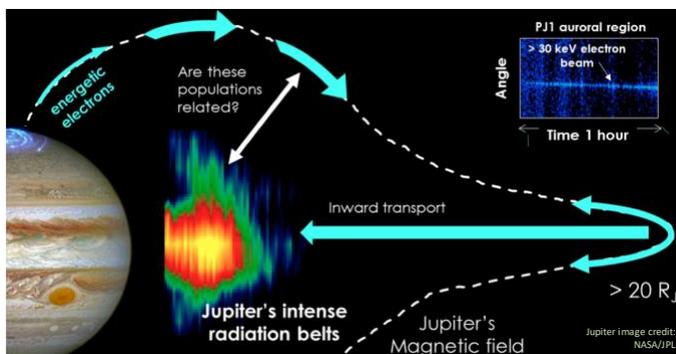

**Figure 5:** The concept of high-latitude electron sources evolving into radiation belt populations

Recent discoveries made by NASA's Jupiter polar orbiting spacecraft, Juno *[Bolton et al., 2017]*, have put the spotlight on auroral processes as a means to generate high-energy ions and electrons that fill Jupiter's plasma sheet. Juno has observed intense and energetic (>1 MeV) electrons beaming upward from the planet *[Mauk et al., 2017a; Clark et al., 2017a]*, energetic ion conics *[Clark et al., 2017b]* and upward ion beams *[Mauk et al., 2018]*. A remarkable discovery is the prevalence of upward electron beams. *Mauk et al. [2017b]* have shown these to be a persistent characteristic of Jupiter's main auroral and polar cap regions. The energy flux contained in the upward beams is often greater than their precipitating counterparts and much greater than for electrons of similar energy in Jupiter's radiation belts. Similarly, in the polar cap region, upward beams of energetic electrons are always present in observations, suggesting active and persistent acceleration at altitudes below the spacecraft *[Mauk et al., 2017b]*. In addition, spectral features of Jupiter's UV aurora in the polar cap observed by Juno, cannot be reconciled with the precipitating electron fluxes *[Ebert et al., 2019]*, hinting that interesting physics occur at lower altitudes that have yet to be explored. Before Juno, *Simpson et al. [1992]* reported >16 MeV electrons and <0.3 keV protons generated at high latitudes with Ulysses data. Juno has detected similar features as bursts of >10 MeV electrons *[Paranicas et al. 2018; Bonfond et al. 2018]*. It is reasonable to hypothesize that a certain fraction of all these upward energetic particles populate Jupiter's plasma sheet and later end up via inward radial diffusion or injections into the radiation belts. Similar hypotheses have been put forward for the MeV electrons produced impulsively over a large region of Saturn's high latitude magnetosphere *[Roussos et al. 2016]*, and for energetic particles from Earth's cusp *[e.g. Frizt, 2001]*, attesting to the possibility of high magnetospheric latitudes being an important radiation belt source, common to strongly magnetized planet with a dense atmosphere.

**Key measurements & justification:** High-latitude auroral acceleration is a constant source of high-flux, field-aligned energetic ions and electrons. The exact details of the particles' fate are unknown and only now are theoretical ideas starting to emerge on their origin *[Elliott et al., 2019]*. It remains unclear if, how and what percentage of the auroral particles convert from field-aligned to trapped, populate the plasma sheet and, subsequently, the radiation belts. The equatorial counterparts of the high-latitude events may be studied partly by JUICE. Still, both JUICE and Juno offer limited information of the highest energy particles as spectral information goes to few MeV. Energy-resolved measurements extending to higher energies would constrain the spectral shape of the high latitude sources, determine if there is an upper limit in the acceleration for both ions and electrons and allow us to compare if the belts' seed population shares common properties with the high latitude accelerated products. Observations in plasma, magnetic field, electric field and waves could be revealing about the origin of the acceleration mechanisms *[McKibben et al., 1993: Clark et al. 2018]*.

## 2.5 The space weather of Jupiter's radiation belts

**State of the art and open questions:** Similar to the magnetosphere that encloses them, Jupiter's radiation belts are very dynamic. This variability, i.e. the radiation belt's space weather, is important to assess for two reasons. Firstly, different physical radiation belt processes which may be difficult to distinguish in average profiles, can be resolved through their distinct variability time scales. Studies for Saturn's proton belt dynamics, for instance, have been central for understanding their CRAND-related origin *[Roussos et al. 2008; Kollmann et al. 2017a; Roussos et al. 2018a]*. The second reason is that there are physical interactions that are responsive to changes in a local radiation belt environment, such as surface sputtering creating transient satellite exospheres *[e.g. Milillo et al. 2016]*.

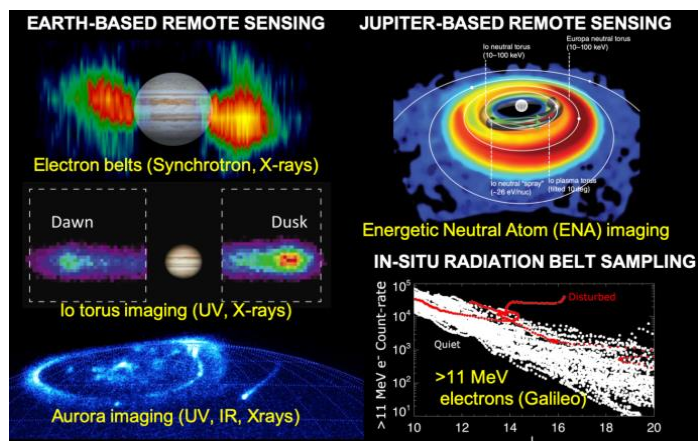

**Figure 6:** The different options for space weather monitoring of Jupiter's radiation belts. Image/plot credits: synchrotron belts (NASA/JPL), torus imaging *[Murakami et al. 2016]*, aurora image (NASA/STScI), ENA image simulation *[Brandt et al. 2018]*, in-situ data plot based on *Roussos et al. [2018b]*.

Most of our information for time variations in Jupiter's radiation belts comes from monitoring the synchrotron emissions by their MeV electrons. These reveal changes at different time scales, ranging from days to years. The longest-term changes seem to have a two-year lagged response to the solar wind *[de Pater et al. 1994; Galopeau et al. 2001; Santos-Costa et al. 2008]*. *Han et al. [2018]* argued that this time lag is explained by slow radial diffusion of electrons mediated through a solar-wind modulated dawn-dusk electric field that exists near Io's orbit *[Murakami et al. 2016]*. If this model is correct, radiation belt transients should evolve the fastest for 10-100 MeV electrons (Figure 2, "corotation resonance"). *Roussos et al. [2018b]* report on such rapid enhancements in >11 MeV electron data by Galileo, but the lack of energy resolution for these measurements, prevents any definite conclusion. How the solar wind transmits changes in the dawn-dusk electric field at Io is also unknown, in part because instantaneous solar wind parameters upstream of Jupiter come from models propagating measurements at 1 AU to the planet's heliocentric distance at 5.2 AU. Furthermore, the dawn-dusk electric field has never been detected in-situ, in plasma flows or electric field measurements *[Bagenal et al. 2016]*. An in-situ detection could verify if this convective fields extends inward or outward of Io's orbit.

Short-term variations in the synchrotron belts (~days) have been observed in response to solar EUV flux changes, supporting a theoretical link between radial diffusion of electrons from changing winds in the planet's thermosphere *[Brice and McDonough, 1973; Tsuchiya et al. 2011; Kita et al. 2013]*. Measurement of electron spectra following EUV changes, for exploiting this link in-situ, are missing. Such measurements could peer into the physics of thermosphere-driven diffusion: Is this process energy dependent and how do diffusion rates scale with L-shell? At Saturn, for instance, where there is evidence that this form of diffusion acts on its radiation belts, in-situ data show that this scaling is much steeper than theory predicts *[Roussos et al. 2007; Kollmann et al. 2017a]*. The least explained variations are those on time scales of weeks, which show no obvious correlation to either EUV or dawn-dusk electric field changes were found *[Kita et al. 2019]*.

For the ion radiation belts of Jupiter, that cannot be accessed through synchrotron emissions, variations are poorly explored. Unlike Saturn, whose moons restrict MeV ion transport, at Jupiter this does not occur below the 40 MeV/n range *[Nénon et al., 2018; Section 2.3]*. Communication between energetic ion populations across the moon orbits should be possible, such that variations in the middle magnetosphere may be transmitted to the innermost belts. For all species, a several day recurrence rate of tail reconnection events, which likely enhance the seed radiation belt population *[Kronberg et al. 2007; Louarn et al. 2014]*, may be important. Juno findings of high latitude particle acceleration sites may also be significant, if these indeed regulate the belt's seed population (Section 2.4). For certain species, the lack of variations may be revealing. For instance, since CRAND (Section 2.2) is a quasi-stable source, seeking for a steady proton signal over multiple orbits would allow to distinguish CRAND from other variable sources of protons. If the population of energetic heavy ions (>40 MeV/n) is stable over even long-time scales, as hinted in data shown in Figure 4B, its source process would need to be equally invariant with time.

**Key measurements & justification:** Any of the aforementioned open questions would benefit from the in-situ measurements identified for all science themes discussed in Sections 2.1-2.4, collected over extended time periods to enable variability studies. Different time scales can be probed by regulating the resampling rate of the radiation belts, as done e.g. for Saturn with Cassini *[Roussos et al. 2014, 2018b]*. Solar wind measurements are important to obtain close to Jupiter. Capturing the global dynamics of the belts' seed population (beyond Europa's orbit), could be achieved through ENA and/or X-ray imaging. Remote sensing of the Io plasma torus, the aurora and the synchrotron emissions of the electron belts would provide excellent context.

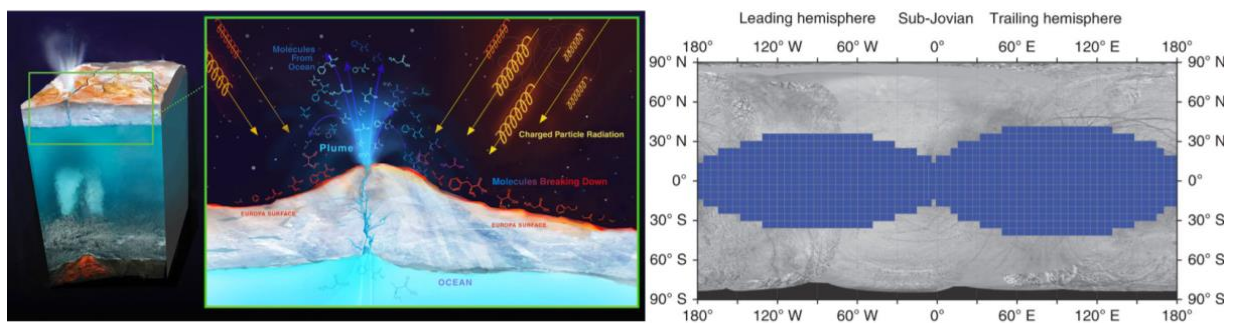

**Figure 7:** (Left) Europa's surface is continuously bombarded by charged particle radiation *(Image credit: NASA/JPL)*. The blue-shaded area (right) shows the current best estimate on where Europa's surface is radiation-processed down to ~10 cm *[Nordheim et al. 2018]*.

## 2.6   Radiation processing of moon surface material and astrobiological implications

**State of the art and open questions:** Excluding Io, the surface of which is regularly altered by its ongoing volcanic activity, all other moons of Jupiter that orbit within the magnetosphere and the radiation belts are exposed to a variety of external weathering processes that modify their physical and chemical properties. Irradiation by low and high energy charged particles is one of these processes. The radiation belt particles, which are also the most surface penetrating, can reach from several cm to several m of depth, either directly or due to the bremsstrahlung they emit as they are decelerated in ice *[Nordheim et al. 2017, 2018]*. Such interactions can transform the top layers of the surface by altering its crystallinity, its thermal inertia or even by creating new molecules *[Paranicas et al. 2018 and references therein]*. These modifications should be considered when interpreting remote sensing observations of the moons (e.g. reflectance spectra), when surface samples are retrieved directly through a lander, or indirectly, by measuring sputtered products and micrometeoroid ejected dust grains from the surface, as it will be done with JUICE and Europa Clipper.

For Europa, where future landers could seek biosignatures in material that may have upwelled from its subsurface ocean, the study of radiation-induced surface modifications has an added value. Particle radiation can impact Europa in different ways, for example by generating the already detected $H_2O_2$ on its surface *[Carlson et al. 1999]*. This oxidant may provide a source of chemical energy to sustain a biosphere in Europa's ocean *[Chyba 2000; Vance et al. 2016]*. On the other hand, *Nordheim et al. [2018]* estimated that aminoacids, the simplest compounds to be sought as potential biosignatures on the surface, can be destroyed by particle radiation from Jupiter down to a depth of about 10 cm in Europa's equatorial region (Figure 7).

Quantifying any of the mechanisms discussed above depends significantly on the input energetic particle spectra to which a surface we assume is exposed. Particularly for >1 MeV electrons and >10 MeV(/n) protons and ions, energy spectra are poorly constrained. Many of the relevant studies rely on arbitrary extrapolations to energies of 100 MeV(/n) or more to predict the most penetrating particles' effects. For even higher energies, it is assumed that the particle spectrum is that of GCRs that can reach into a moon's orbit *[Nordheim et al. 2018]*, resulting in rather low input fluxes for most moons: for Europa, <13 GeV GCRs cannot reach its orbit *[Selesnick, 2002; Nordheim et al. 2019]*. However, if processes like CRAND provide even a small input to Jupiter's radiation belts (Section 2.2), fluxes of relativistic protons can be significantly elevated from their presumed "empty" GCR levels. Proton trapping at Europa is limited at ~2.5 GeV and exceeds 100 GeV near the planet *[Birmingham 1982]*. CRAND protons would populate the spectral range below the GCR cutoff energy, up to the trapping limit.

**Key measurements & justification:** An improved description of the energetic particle spectra would transform our understanding of how Jupiter's radiation belt environment affects the surfaces of its moons. All effects discussed here develop on time scales of thousands to millions of years, such that long-term average energetic particle spectra should be sufficient to estimate the energy flux and its long-term impact on the moons' visible surfaces and subsurfaces. The requirements for the energy range that the measurements should cover are similar to those discussed in Sections 2.1-2.5. Understanding if certain physical processes, like CRAND, contribute to the radiation belt content, would allow to physically extrapolate measurements to even higher energies than it would be possible to detect.

## 3  SPACE MISSIONS TO JUPITER'S RADIATION BELTS

Here we outline considerations for space missions that could explore Jupiter's radiation belts and we identify enabling technologies. Several past studies by ESA were done in the context of the JUICE mission, including its previous variants, Laplace and EJSM. Even earlier studies of relevance are those of the Jupiter Radiation Study (JURA) and a series of investigations for the exploration the jovian system by *Atzei et al. [2007]*. The latter were subdivided into three mission studies, the Jovian Minisat Explorer (JME), the Jupiter Entry Probe (JEP) and the Jovian System Explorer (JSE), focusing on Europa, the jovian atmosphere and the magnetosphere, respectively. The JUpiter MagnetosPheric boundary ExploreR (JUMPER) study is likewise relevant *[Ebert et al. 2018]*. Appropriate points from those reports are reiterated here. Besides that, the mission scenarios discussed in Section 3.3, which are only introduced as concepts, deserve dedicated studies.

### 3.1  Scientific design drivers

**Spatial coverage:** Studying the radiation belts requires to monitor them as part of the jovian magnetosphere. Spatial coverage should be maximized in terms of distance (from the planet to the magnetopause), latitude and local time, similar to what Cassini achieved in Saturn's magnetosphere. Should that not be possible, the least explored regions should be prioritized, i.e. the radiation belts' core between the planet and Io at low and mid-latitudes. In terms of local time, the strongest asymmetries are expected between the dawn and dusk sectors.

**Temporal coverage:** The radiation belts evolve on a variety of time scales, the longest ones in the order of few years. The minimum lifetime of a single mission to Jupiter's belts should be about 2 years. Resolving shorter time scales (weeks) could be achieved through radiation belt crossings with a similar frequency. For even shorter time scales and for separating temporal from spatial variations, multi-point, in-situ measurements in the magnetosphere and the synergy between remote sensing and in-situ measurements are important. Remote sensing is offered through Earth based observatories (e.g. synchrotron observations – Table 1, Figure 6) or from Jupiter, at a wide range of wavelengths as well as in ENAs.

**Instrumentation:** While the radiation belts are described by their energetic particle distributions, their origin, structure and dynamics are best understood through a comprehensive investigation of their space environment in plasma, magnetic, electric field and waves at a wide frequency range (Sections 1 and 2). Table 3 marks which type of instruments are relevant for different investigations. The properties for the particles and fields suite of Juno and JUICE *[Bolton et al. 2007; Grasset et al. 2013]*, offer excellent analogues for the requirements of similar detectors in a mission to the radiation belts. An added capability should be that of energetic ion charge states, as done with the CHEMS instrument on *Cassini [Krimigis et al. 2004]*. Among the instrument types listed in Table 3, the most novel would be those for detecting (ultra)relativistic particles. Extending the measurement capabilities to ~100 MeV electrons and ~1 GeV(/n) ions, in order to comply with expectations for the energies of trapped populations in Jupiter's radiation belts, has never been attempted. Measurements of particle distributions for all energies requires 4π sky coverage for ions or full pitch angle coverage for electrons, such that a spinning spacecraft is preferred over a three-axis stabilized one. A spinning spacecraft may have additional advantages for instrument design (Section 3.4).

| | Plasma | Magnetic field | Electric field | Waves | Energetic particles | | ENA and/or X-Ray imaging (from Jupiter) | Solar wind in-situ monitoring | Earth-based monitoring |
|---|---|---|---|---|---|---|---|---|---|
| | | | | | Non-relativistic | Relativistic | | | |
| Electron radiation belts (2.1) | X | X | X | X | X | X | X | X | X |
| CRAND protons (2.2) | | | | | X | X | | | |
| Ion radiation belts (2.3) | X | X | X | X | X | | X | X | X |
| High latitude sources (2.4) | X | X | X | X | X | X | X | X | X |
| Space weather (2.5) | X | X | X | X | X | X | X | X | X |
| Moon surfaces (2.6) | | | | | X | X | | | |

**Table 3:** Science questions (Section 2), relevant instruments and observation modes.

### 3.2 Technical design drivers

While there are numerous design drivers, below we focus on those with unique aspects for a mission to Jupiter's radiation belts (mass, radiation, communications, planetary protection). Discussion about power, thermal constraints, spacecraft autonomy etc. are adequately addressed in past Jupiter mission studies.

**Mass:** The payload masses for Juno, JUICE and Europa Clipper are ~170, ~220 and ~350 kg respectively *[Bolton et al 2007; ESA, 2015; NASA, 2018]*, with typically less than 40% allocated for in-situ particles and fields instrumentation and ENA imagers. While that appears to imply significant extra mass for accommodating improved and new instruments for a dedicated space physics payload, the available margin may be significantly smaller: The required, multiple crossings through the radiation belts translate to heavier radiation shielding. The possible use of instruments with strong magnets may also require additional material for magnetic shielding (Section 3.4). The need for solar wind monitoring and multi-point in-situ measurements implies mission designs with at least two orbiting spacecraft, setting tight mass constraints for the case these have to be accommodated in a single launcher. From the studies of *Ebert et al. [2018]* and *Atzei et al. [2007]*, the minimum mass of a second orbiter would be in the range of 100 kg. Payload distribution among the different spacecraft, orbit, instrument and shielding design etc. are parameters that need to be iterated for maximizing payload mass.

**Radiation:** For each of its few orbits through the radiation belts' core the Galileo spacecraft was accumulating 30-40 krad per crossing behind an equivalent shielding of 2.2 g/cm$^2$ aluminum. This means that within ~9 orbits, a radiation belt crossing spacecraft would accumulate an equivalent dose as the Galileo spacecraft during its 6.5 years and 34 orbits mission *[Fieseler et al., 2002; Atwell et al., 2005]*. Measures should thus be taken, through a combination of orbit design, spacecraft/instrument shielding, instrument accommodation and the use/development of radiation-hard subsystems, to minimize radiation dose accumulation per orbit, to extend subsystem radiation resistance and overall maximize mission lifetime and the number of radiation belt crossings.

**Communications:** In the case of a multi-spacecraft mission the options for autonomous (Earth-direct) vs intra-spacecraft (relay) communications should be examined as they impact orbit and spacecraft design. Direct-to-Earth communications likely allow for higher flexibility and autonomy to each spacecraft for science operations. A relay offers the option of longer storage and downlink periods of large volume of high-quality data *[Atzei et al. 2007]*.

**Planetary protection:** Europa's surface or immediate subsurface (see Section 2.6) may be hosting biosignatures associated to the subsurface ocean below its crust, and thus planetary protection considerations are relevant, especially for a mission that may cross the moon's orbit and a spacecraft that is subject to risks imposed by very high radiation exposure. In terms of orbit, the design should minimize the possibility of an accidental impact with the moon. In the study for the Jovian Minisat Explorer it was considered that in-flight decontamination by the extreme radiation in Jupiter's belts may be an option that could be explored.

## 3.3   Mission considerations

Monitoring of the radiation belts through one or more orbiters is necessary for their comprehensive exploration. Two different orbiter mission concepts are outlined, each one for different mission cost levels (L- & M-class). A third flyby mission option is described as a target-of-opportunity or a pre-cursor to options (1) and (2) (Figure 8). Any of the three scenarios would rely on coordinated Earth-based monitoring of the jovian system (Table 1). For any of the mission concepts, international collaboration may reduce the ESA's contribution and depending on the level of involvement the cost-category of a mission scenario may change. All concepts are feasible on solar power, although other options involving space nuclear power *[e.g. Tinsley et al. 2017]* would allow for extra flexibility in mission design.

**1)     L-Class (Radiation belt orbiter & Solar wind/magnetosphere monitor):** Covering all science goals identified in Section 2 requires synergistic investigations by minimum two orbiters. For the minimal configuration we consider that a primary spacecraft performing regular radiation belt crossings carries the in-situ particles and fields instrumentation suite, while a smaller spacecraft monitors the upstream solar wind and the magnetosphere globally.

The primary spacecraft's orbit should be highly elliptical, similar to Juno's (~53 days period), to minimize crossing durations through the belts and exposure to radiation, but more equatorial so as to sample the radiation belts at their core. As both Juno and Cassini investigations already demonstrated, belt crossings lasting several hours each are more than sufficient for collecting excellent quality observations and for resolving radiation belt structuring down to ~$10^{-2}$ planetary radii thanks to the high-time resolution sampling that is possible for all particles and fields instruments *[Kollmann et al. 2017b; Roussos et al. 2018a]*. As the mission evolves, the eccentricity and the period of the orbits could be reduced so as to achieve more frequent belt crossings and capture shorter variability time scales. Excursions in local time and latitude could be performed as the apojove reduces, e.g. through Callisto flybys, like it will be done for JUICE, or with spacecraft resources, if possible. Latitudinal excursions could regulate the rate of ionizing dose accumulation as the belt crossings are becoming more and more frequent.

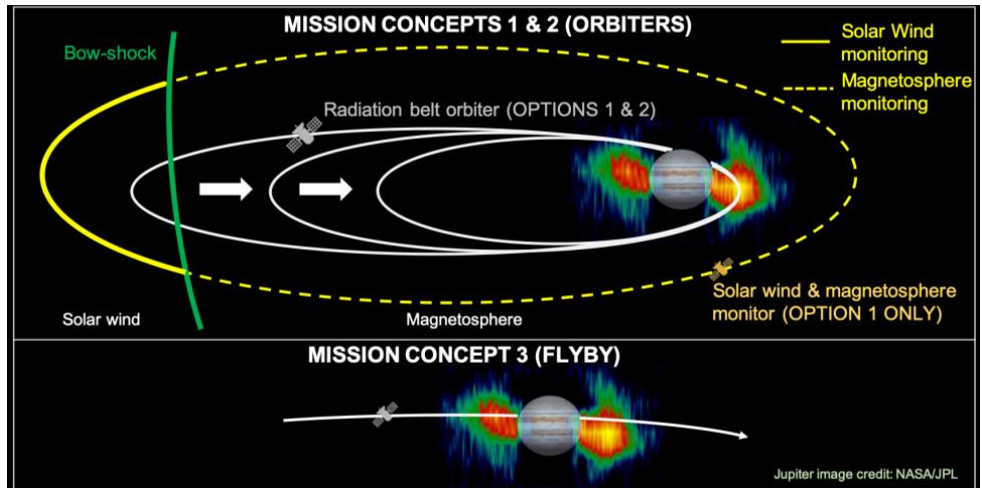

**Figure 8:** Simplified sketch for orbiter mission concepts "1" and "2" (top). Option "3" (flyby) is shown at the bottom panel. Distances shown are not to scale.

The secondary spacecraft, carrying a lightweight in-situ particles and fields suite and remote sensing instruments (e.g. *Ebert et al. [2018]*), should be placed in an orbit extending upstream of Jupiter's bow-shock (~100 $R_J$) from where solar wind monitoring is possible for extended periods. Its perijove should be outside Callisto's orbit to limit radiation exposure. Closer to the magnetosphere, ENA and X-Ray imaging becomes possible, along with simultaneous two-point measurements with the primary orbiter. If only one imager can be accommodated, an ENA camera would be preferred if all the other remote sensing methods that can be executed from Earth (Table 1) are available. X-ray imagers offer otherwise excellent remote sensing options, as they may probe simultaneously the electron belts, the aurora and the Io torus *[Elsner et al. 2002; Dunn et al. 2017; Branduardi-Raymont et al. 2018; Numazawa et al. 2019]*.

2) **M-class (Radiation belt orbiter):** The M-class variant is similar to the L-class but does not include the solar wind/magnetosphere monitoring spacecraft. Most science goals that depend on solar wind monitoring, particularly 2.5 but also 2.1, 2.3 and 2.4, are affected. Losses can be mitigated by propagating solar wind parameters from 1 AU. Close-proximity ENA or X-Ray imaging and two point in-situ measurements in the magnetosphere cannot be replicated. The advantages of a single orbiter are that mission design and operations would be simpler, while significant mass allocation could become available for advanced instrument design, radiation shielding or other critical mission elements. This variant retains a great discovery potential, given that Jupiter's radiation belts are poorly explored in-situ (Section 1.2).

3) **F- or S-class (Flyby mission):** Due to its strong gravity, Jupiter is regularly used as a swing-by target for adjusting a spacecraft's trajectory towards its solar system destination. Many mission concepts target the Jupiter system itself. A CubeSat or Smallsat-sized probe riding along with the main spacecraft could be released for a flyby through the belts' core. For such a mission, observations should prioritize measuring the least defined parameters, i.e. suprathermal ion charge states, electrons up to ~100 MeV, protons up to ~1 GeV and ion composition at least up to 100 MeV/n. Observations from such a flyby mission could provide key input for science goals 2.2, 2.3 and 2.6, further constraints to all others and refine the requirements for mission options 1 and 2. Should resources allow, a small, short-lived orbiter can be considered instead: Such low resource missions can be successful: data from the Colorado Student Space Weather Experiment CubeSat in Earth's radiation belts, for instance, led to the discovery of the decades-sought, trapped CRAND electrons *[Li et al. 2017]*.

## 3.4 Enabling technologies

**Miniaturization of relativistic charged particle detectors:** For the high energies considered in Jupiter's radiation belts (~100 MeV electrons, ~GeV/(n) ions), particles are both penetrating and fast. Methods measuring time-of-flight or energy losses on silicon detector stacks either don't work or give poor constraints to a particle's energy, e.g. above 100-200 MeV for protons. A most efficient method to resolve energy, species and charge state of such particles is magnetic spectrometry. Existing detectors that use it are very power consuming and heavy (many 10s of kg or more) partly due to the application of large permanent magnets *[e.g. Picozza et al. 2007]* and partly because they are usually designed to detect ultra-high energy GCRs. A step towards miniaturizing such instruments is the MiniPAN (Mini-Penetrating Particle Analyzer) detector *[Wu et al. 2018, 2019]*, a modular magnetic spectrometer (mass <10 kg, power <10 W) that may perform ion composition measurements up to several GeV(/n). A prototype was approved for development through a European Union Future and Emerging Technologies program, but a Jupiter-specific design may require further studies. The energy limit to which magnetic spectrometry can be applied for resolving heavy ion charge states remains to be investigated. Non-magnetic spectrometers for ultrarelativistic electrons developed in Europe, such as HEPD that is operating on the China Seismo-Electromagnetic Satellite *[Alfonsi et al. 2017]*, use silicon detector stacks and calorimeters. Their large mass can be greatly reduced by adjusting their geometry factor to the orders of magnitude stronger >1 MeV electron fluxes at Jupiter (Figure 1). Additionally, such instruments reconstruct the 3D trajectory of the measured particles by tracking them on large, successive detector planes. With an instrument on a spinning spacecraft (Section 3.1) tracking becomes 2D, reducing the total mass and power consumption as planar detectors can be significantly decreased in size.

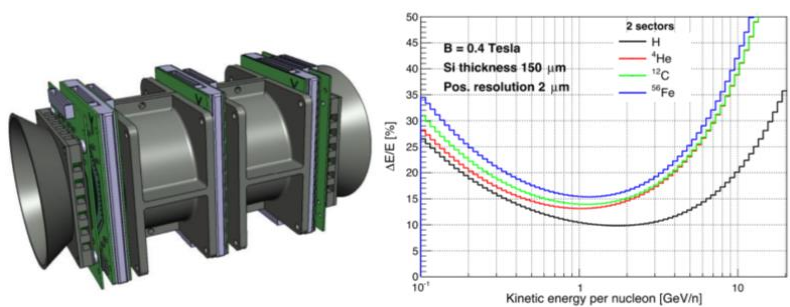

**Figure 9:** (left) Simplified schematic of a double-sided MiniPAN instrument concept. (right) The estimated energy resolution as a function of ion kinetic energy, for different ion species *[Wu et al., 2018, 2019]*.

**Electromagnetic cleanliness:** For magnetic spectrometry to capture the energies particles in Jupiter's radiation belts, permanent magnets should generate static fields about 2-5 times stronger than what has been used e.g. on the LEMMS instruments of Cassini and Galileo (~0.06-0.08 T) *[Williams et al. 1992; Krimigis et al. 2004]* or the MagEIS instrument of the Van Allen probes (~0.15 T) *[Blake et al. 2013]*. Strong stray fields from permanent magnets can interfere with plasma detectors or magnetometers. The trade-off between the maximum energy resolved (determined by the magnet strength and size), the strength of the stray fields and instrument mass should be considered. Instrument accommodation and design, e.g. choice of materials for the permanent magnet and for magnetic shielding, play a role.

**Radiation hard electronics and detector performance:** The JUICE mission is designed such that the Total Ionizing Dose (TID) experienced by its most sensitive electronics contained in heavily shielded vaults would not exceed 50 krad *[ESA, 2015]*. This tolerance level should be significantly increased for a dedicated radiation belt mission. In addition, developing efficient active shielding systems for particle detectors (e.g. anticoincidence or positive coincidence) that would reject counts from high fluxes of penetrating MeV electrons and protons in Jupiter's belts, should be investigated. Low energy electron detectors (<10 keV) are the most challenging for using anticoincidence *[e.g. McComas et al. 2017]*, as techniques like Time-of-Flight, cannot be applied, and positive coincidence schemes should be explored. Both the high-resistant electronics and active shielding techniques, would reduce the requirements for passive spacecraft shielding as the way to achieve a good signal to noise ratio, allowing to save mass or reallocate it in other subsystems.

**Energetic Neutral Atom imaging (>>1 keV):** ENA imaging is considered in the strawman payload described in Section 3.1. JUICE will be first mission to perform a dedicated study at Jupiter in low and high energy ENAs *[Grasset et al. 2013]*. Still, detectors for >>1 keV ENAs (INCA on Cassini, JENI on JUICE), have been or are being developed only outside Europe. Developing high energy ENA capabilities would enable a Jupiter radiation belts mission to proceed independently of contributions from international partners, and open new possibilities for scientific applications in Earth's magnetosphere, comets and the heliosphere where ENA imaging is applicable *[McComas et al. 2009; Dialynas et al. 2017; Brandt et al. 2018]*.

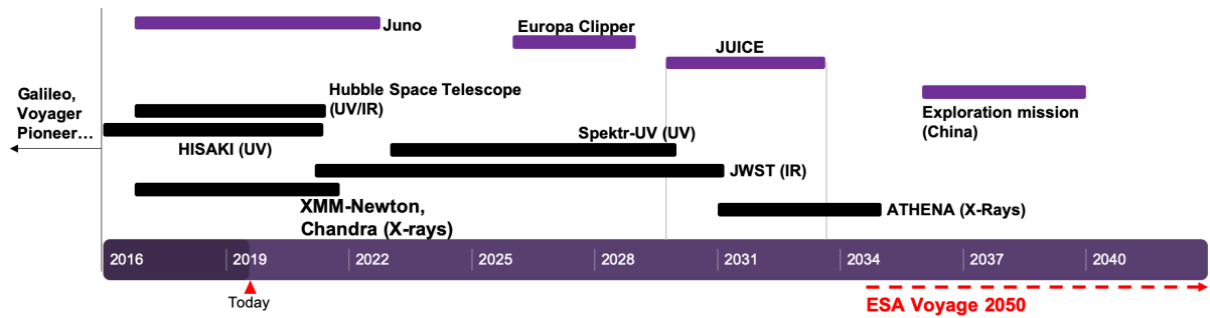

**Figure 10:** The timeline of missions relevant to Jupiter's magnetosphere and radiation belts (magenta: space missions, black: Earth-based observatories that can target Jupiter). Ground-based, synchrotron radiation belt or IR, jovian aurora observations are assumed to be available throughout this timeline.

## 4 JUPITER RADIATION BELT SCIENCE IN CONTEXT

### 4.1 Jupiter exploration

Following the exciting discoveries by the Galileo mission at Jupiter, JUICE and Europa Clipper were designed in response. Besides exploring Europa, Ganymede and Callisto in detail, these missions will carry instruments that offer an interdisciplinary view of the jovian system, including its magnetosphere and outer radiation belts. Similarly, important discoveries by Juno, which has been extended through 2022, are revealing the complexity of the jovian magnetosphere and its energetic particle environment, already generating many new questions (Section 2.4). These missions, along with campaigns by Earth-based observatories, would shape our views about Jupiter until about 2035 (Figure 10). Beyond 2035, a mission that may explore Jupiter's magnetospheric environment has been announced by China, with arrival probably after 2036, but with no further details known *[CAS, 2019]*. A space mission, picking up from where Juno, JUICE and Europa Clipper take our understanding, will continue the exploration of the giant planet, which has been regular since Pioneer 10 and 11 (Table 1). Jupiter exploration has also benefited from the synergy of space missions and Earth-based observatories. The prospects for the development of new observatories after 2035 are still unclear although already being considered *[e.g. Murakami et al. 2019]*.

### 4.2 Multi-point investigations of planetary magnetospheres

Over the last two decades, terrestrial magnetospheric science has benefited greatly from space missions utilizing several spacecraft for simultaneous, multi-point measurements of the geospace, in addition to the regular monitoring of the upstream solar wind and of the various geomagnetic indices which has been going on for even longer. Missions like Cluster, Double Star, THEMIS, Magnetospheric Multi-Scale (MMS) and the Van Allen Probes have drastically changed the way we understand Earth's radiation belts, as part of the terrestrial magnetosphere (e.g. *Escoubet et al. [2015]*). The findings from the Van Allen Probes, in particular, have led to many breakthrough discoveries and have evolved the state of the art of radiation belts' research to a new level *[Shprits et al. 2016; Ozeke et al. 2017; Baker et al. 2018; Selesnick and Albert, 2019]*. The remarkable increase of refereed publications on the subject after 2000 (Figure 11) and the launch of follow-up missions like Arase, the China Seismo-Electromagnetic Satellite, Lomonosov and others [*Alfonsi et al. 2017; Shprits et al. 2018b; Miyoshi et al. 2017*], are representative of the aforementioned, rapid progress and increased interest in the field. Multi-point measurements of space plasma interactions in our solar system tend to become a norm: The two ARTEMIS spacecraft investigate the lunar-solar wind interaction since 2010 *[Halekas et al. 2011]*, while Bepi Colombo will do the same at Mercury *[Benkhoff et al. 2010]*. The recently funded ESCAPADE mission will be the first dedicated two-orbiter mission to study the Mars-solar wind interaction *[Lillis et al. 2018]*. Similarly, ESA's Comet Interceptor will perform two-point measurements of a comet's plasma environment *[Jones et al. 2018]*. As it stands now, no coordinated multi-spacecraft mission is planned for any outer planet magnetosphere, but it is the essential and obvious way forward. A mission to explore Jupiter's radiation belts (Section 3.2) would have an enormous potential for discoveries, not only because it could be the first to investigate Jupiter's magnetosphere with multiple spacecraft, but also because it would focus on its least explored component.

### 4.3   Comparative studies

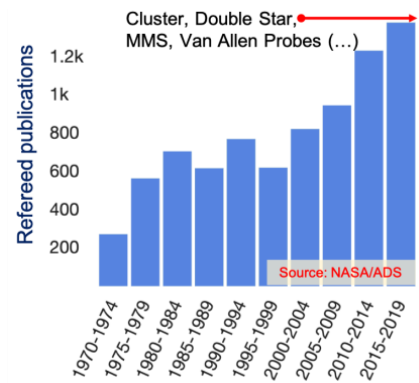

**Figure 11:** Number of refereed publications on energetic charged particles and/or the radiation belts of Earth between 1970 and 2019 from https://ui.adsabs.harvard.edu/[1]

Jupiter's radiation belt science is well aligned with the continuous investigations of Earth's radiation belts by numerous ongoing missions, and the fact that many White Papers in response to the Voyage 2050 call focus on charged particle acceleration and energy transfer between plasmas and fields (e.g. *Rae et al. [2019]*). Thanks to the extensive exploration of Saturn's radiation belts by Cassini and the continuously growing set of measurements by Juno at Jupiter, a large volume of high-quality observations is being generated covering many years and different radiation belts, enabling detailed comparative studies for the first time. Planned or ongoing missions would add to that: the limits of radiation belt formation can be exploited with Bepi-Colombo at the weakly magnetized Mercury, and JUICE at Ganymede's mini-magnetosphere *[Eviatar et al. 2000]*. Potential extension of such observations to Uranus or Neptune *[Fletcher et al. 2019]* would allow us to explore radiation belts across a variety of environments and scales. X-ray imagers, following their application on SMILE and Bepi-Colombo *[Benkhoff et al. 2010; Branduardi-Raymont et al. 2018]*, will become readily available for Jupiter missions and open a new are in the exploration of planetary magnetospheres.

### 4.4   Opportunities for additional and interdisciplinary science investigations

**Magnetospheric and Jupiter system science:** A mission designed for radiation belt studies is suited to explore any other magnetospheric region it crosses. The mission concepts considered (Section 3.3) are sufficient to extend and/or complement many of the magnetospheric science goals of Galileo, Juno, JUICE and Europa Clipper *[Grasset et al., 2013; Pappalardo et al. 2017; Bagenal et al. 2017]*. The option for coordinated, multi-point measurements could be a first for a giant planet's magnetosphere, with obvious advantages (Section 4.2). The strong coupling between the radiation belts and Jupiter offers many opportunities to study different components of the jovian system, besides what is described in Section 2.6. For instance, Juno data show that Jupiter's internally generated magnetic field has changed in the time between Pioneer 10/11 observations and today *[Connerney et al. 2017; Moore et al. 2019]*. The observed variations, attributed to advection of the field by Jupiter's zonal winds, could be further monitored with a radiation belt mission reaching very close to Jupiter two decades after Juno. Potential close moon flybys offer a chance to conduct

---

[1] Generated with the command: abs:( ((radiation belt) OR (energetic particles)) NOT Jupiter NOT Saturn NOT Uranus NOT Mars NOT Mercury NOT Neptune AND Earth) AND year:1970-2019

new type of measurements: the detection of high energy albedo particles from moon surfaces, released due to the precipitation of very high energy ions is an example. Such particles, recently detected at Earth's moon, offer insights on the physical properties of the immediate subsurface [*Wilson et al. 2012; Schwadron et al. 2016*].

**Heliospheric physics:** ENA imaging offers opportunities to explore the formation, interactions and the large-scale structure of the global heliosphere, providing insights on the plasma processes at ~100 AU and beyond. Relevant investigations by the Interstellar Boundary Explorer (IBEX) and Cassini/INCA have already revolutionized our notions about the global heliosphere [*Krimigis et al. 2009; McComas et al. 2009; Dialynas et al. 2017*], established an increasingly growing community and new missions either in preparation (IMAP [*McComas et al. 2018*]) or in consideration (Interstellar Probe [*McNutt et al. 2019*]). Potential measurements by an ENA imager at Jupiter would extend the time history of previous measurements. Regular GCR observations and a detailed view of escaping relativistic electrons from Jupiter, offered by the mission's energetic particle detectors, when combined with similar observations at 1 AU, would update models of particle transport in the heliosphere [*Vogt et al. 2018*].

**Astrophysics:** Jupiter offers many astrophysical parallels (Section 1.3). Understanding the processes at work in Jupiter's radiation belt can for instance inform the study of rotating radio transients which may originate from particle scattering within the pulsar radiation belts [*Luo and Melrose, 2007*]. In-situ measurements offer insights on how a strong charged particle accelerator can seed its surrounding environment with high energy particles. The study of Jovian Cosmic Rays, i.e. energetic electrons and ions escaping into the heliosphere [*Marhavilas et al. 2001; Vogt et al. 2018*], in conjunction with similar studies at other planets [*Krimigis et al. 2009; Mauk et al. 2019*] could update models of how this escape develops and which parameters control it, with implications about "local cosmic ray" sources in different astrophysical systems.

**Radiation belt environment models:** Jupiter's mini-solar system is a very attractive target for many exploration missions. For any of these, considerations of Jupiter's radiation environment are always a high priority. Radiation environment models, bound to the existing data limitations, have uncertainties (e.g. *Evans et al. [2013]*) that could propagate into mission design in the form of extra shielding mass (or less payload mass). A comprehensive in-situ investigation of Jupiter's radiation belts would support empirical radiation belt model design and offer opportunities to optimize the future exploration of the jovian system and many of its rings and moons that orbit within its harshest energetic particle environment.

## 5   SUMMARY

Jupiter's radiation belts are a major component of the planet's magnetosphere, but the unprecedented scale, complexity and energetics make their measurement very demanding. Their severe energetic particle environment has challenged, and still challenges, their in-situ spacecraft exploration. However, the outstanding science that can be performed through their in-situ study should motivate us to overcome these hurdles and plan for their comprehensive survey. The Juno mission has passed repeatedly through Jupiter's belts and has shown this to be a feasible technical task. Experience gained from the JUICE mission, and the highly anticipated results of its scientific investigations, will pave the way for the next step in the exploration of the jovian system by ESA. In this White Paper we have argued why a multi-spacecraft mission to Jupiter's radiation belts should be that next step. A first multi-spacecraft mission to a giant planet's magnetosphere and radiation belts lends itself to numerous important discoveries, as the extraordinary success of ESA's Cluster mission, a similar "first", has proven. The interdisciplinary character of the investigations that can be performed in Jupiter's mini solar system will only increase the mission's impact further. We therefore believe that Jupiter's radiation belts deserve to be assigned as a high priority target in the Voyage 2050 programme cycle and we urge ESA to conduct and support the necessary mission, instrument and technology studies that would eventually make their in-situ exploration a reality.

## PROPOSING TEAM

1. **Elias Roussos** [Max Planck Institute for Solar System Research, Goettingen, Germany]
2. **Oliver Allanson** [University of Reading, Reading, Berkshire, UK]
3. **Nicolas André** [Research Institute for Astrophysics and Planetology, Toulouse, France]
4. **Bruna Bertucci** [University of Perugia, Italy]
5. **Graziella Branduardi-Raymont** [UCL Mullard Space Science Laboratory, UK]
6. **George Clark** [Johns Hopkins Applied Physics Laboratory, Laurel, Maryland, USA]
7. **Kostantinos Dialynas** [Academy of Athens, Athens, Greece]
8. **Iannis Dandouras** [Research Inst. for Astrophysics and Planetology Toulouse, France]
9. **Ravindra Desai** [Imperial College London, London, UK]
10. **Yoshifumi Futaana** [Swedish Institute for Space Physics (IRF), Kiruna, Sweden]
11. **Matina Gkioulidou** [Johns Hopkins Applied Physics Laboratory, Laurel, Maryland, USA]
12. **Geraint Jones** [UCL Mullard Space Science Laboratory, UK]
13. **Peter Kollmann** [Johns Hopkins Applied Physics Laboratory, Laurel, Maryland, USA]
14. **Anna Kotova** [Research Institute for Astrophysics and Planetology, Toulouse, France]
15. **Elena Kronberg** [University of Munich, Germany]
16. **Norbert Krupp** [Max Planck Institute for Solar System Research, Goettingen, Germany]
17. **Go Murakami** [Institute of Space and Astronautical Science, JAXA, Kanagawa, Japan]
18. **Quentin Nénon** [Space Sciences Laboratory, University of California at Berkeley, USA]
19. **Tom Nordheim** [Jet Propulsion Laboratory, Pasadena, USA]
20. **Benjamin Palmaerts** [University of Liege, Liege, Belgium]
21. **Christina Plainaki** [Agenzia Spaziale Italiana (ASI), Rome, Italy]
22. **Jonathan Rae** [University College London, London, UK]
23. **Daniel Santos-Costa** [Southwest Research Institute, USA]
24. **Theodore Sarris** [Democritus University of Thrace, Greece]
25. **Yuri Shprits** [Helmholtz-Zentrum, Potsdam, Germany]
26. **Ali Sulaiman** [University of Iowa, Iowa, USA]
27. **Emma Woodfield** [British Antarctic Survey, Cambridge, UK]
28. **Xin Wu** [University of Geneva, Switzerland]
29. **Zhonghua Yao** [University of Liege, Liege, Belgium]